\documentclass[twocolumn,secnumarabic,amssymb,nobibnotes,aps,longbibliography,pra,showpacs]{revtex4-1}

\usepackage{bm}
\usepackage{amsfonts,fancyhdr,epstopdf,times}
\usepackage{epsfig}
\usepackage{amsfonts,subfigure}
\usepackage{dcolumn}
\usepackage{bm}
\usepackage{graphicx,times,psfrag,color,hyperref,xcolor}
\usepackage{amssymb,latexsym,mathrsfs,amsmath,amsthm}
\usepackage[utf8]{inputenc}
\usepackage[mathlines]{lineno}
\usepackage{verbatim}
\usepackage{lipsum}
\usepackage[percent]{overpic}

\begin{document}

\title{Noise-robust quantum sensing via optimal multi-probe spectroscopy}

\author{Matthias M. Müller}
\affiliation{\mbox{Department of Physics and Astronomy}, \mbox{LENS}, and \mbox{QSTAR}, University of Florence, via G. Sansone 1, I-50019 Sesto Fiorentino, Italy.}

\author{Stefano Gherardini}
\affiliation{\mbox{Department of Physics and Astronomy}, \mbox{LENS}, and \mbox{QSTAR}, University of Florence, via G. Sansone 1, I-50019 Sesto Fiorentino, Italy.}

\author{Filippo Caruso}
\affiliation{\mbox{Department of Physics and Astronomy}, \mbox{LENS}, and \mbox{QSTAR}, University of Florence, via G. Sansone 1, I-50019 Sesto Fiorentino, Italy.}

\begin{abstract}
The dynamics of quantum systems are unavoidably influenced by their environment and in turn observing a quantum system (probe) can allow one to measure its environment: Measurements and controlled manipulation of the probe such as dynamical decoupling sequences as an extension of the Ramsey interference measurement allow to spectrally resolve a noise field coupled to the probe. Here, we introduce fast and robust estimation strategies for the characterization of the spectral properties of classical and quantum dephasing environments. These strategies are based on filter function orthogonalization, optimal control filters maximizing the relevant Fisher Information and multi-qubit entanglement. We investigate and quantify the robustness of the schemes under different types of noise such as finite-precision measurements, dephasing of the probe, spectral leakage and slow temporal fluctuations of the spectrum.
\end{abstract}

\maketitle

\section{Introduction}

The dynamics of a quantum system is inevitably affected by its usually deliterious interaction with the environment~\cite{Petruccione2003,Caruso2014}. Especially, the advent of quantum technologies, such as quantum computing~\cite{Nielsen1}, quantum communication~\cite{Caruso2014}, and quantum metrology~\cite{Giovannetti2004}
has lead to increasing interest in measuring the environment: indeed, quantum sensing~\cite{Giovannetti2004,Giovannetti2011,Degen2016} itself is considered to be a quantum technology.

Robust control of a quantum system is crucial to perform quantum information processing~\cite{Green2013,Kallush2014,Pawela2015}, which has to be protected from decoherence or noise contributions originating from the environment, employing strategies such as dynamical decoupling~\cite{Viola1999,Khodjastesh2005} and optimal control theory~\cite{Brif2010}. Yet, unfortunately, it can be never completely shielded~\cite{Zhang2007,Kallush2014,Lloyd2014,vanderSar2012}. Indeed, the decay of the coherence depends in a characteristic way on the spectrum of the bath and the control of the system and can be described by a univeral formula \cite{KofmanNAT2000,KofmanPRL2001,Gordon}. On the one hand, this knowledge can be used to exploit the bath to perform tasks such as optimized state transfer \cite{Zwick2014,Caruso2010}. On the other hand, it allows to examine the influence of a specific frequency window of the bath modes by designing a suitable control. The application of different control functions lies at the core of the so-called filter function approach to spectrally-resolved quantum sensing \cite{KofmanPRL2001,Gordon,Degen2016, Paz-Silva2014, Norris2016}. Several protocols for this type of frequency-resolved measurement of the spectral density have been designed based on multiple pulse sequences~\cite{Khodjastesh2005,Cywinski,Uhrig2008,Biercuk2011,Yuge2011,Alvarez2011,Bylander2011,Kotler}.

Experimentally, quantum sensing has been performed with optimal control theory (open-loop) pulses for improved magnetometry \cite{Haeberle2013,Nobauer2015}, and real-time adaptive measurements \cite{Bonato2015} with single-spin sensors. These sensors can be also implemented on scanning device to gain spatial resolution of the field strength \cite{Hall2009}. Noise spectroscopy methods have been tested with a variety of different platforms as super-conducting flux qubits \cite{Bylander2011}, single nuclear spins in polycrystals \cite{Alvarez2011} and NV-centers \cite{Bar-Gill} and have impact also in biomedicine, as demostrated e.g. for desease-detection in mouse brains~\cite{Smith2012,Alvarez2014}.

Recent developments in the field of quantum sensing with temporal resolution include the study of multi spectra \cite{Norris2016,Paz-Silva2017}, which provide more information on the non-Gaussianity of a signal. These studies build on a general transfer-function approach to noise filtering via open-loop quantum control that has been recently introduced in Ref.~\cite{Paz-Silva2014} for Gaussian dephasing environments, and extended to non-Gaussian environments and multi-spectra in Ref.~\cite{Norris2016} and to the use of multiple qubits in Ref.~\cite{Paz-Silva2017}. Ref.~\cite{Zwick2016} introduces a way to determine the main features of an environment by drastically reducing the measurement complexity. To this aim, the spectrum of the environment is parametrized so that the width of the spectrum can be determined by a single measurement, where the precision of this measurement is optimized by help of the Fisher information associated with the measurement and the parameter. Furthermore, the problem of spectral leakage has gained interest~\cite{Norris2016,Wu2009,Frey2017}. Most protocols can investigate the signal only in a finite frequency band, while the interaction of the probe with the environment has contributions also outside this band and those contributions can have a deleterious effect on the measurement precision. This problem has been addressed by parametrizing the high-frequency contributions \cite{Alvarez2011}, as well as by extending the measurement range associated with a given temporal resolution of the control \cite{Norris2016} in the context of dynamical decoupling or bang-bang control~\cite{Viola1999}, while for continuous pulse modulation \cite{Gordon,Kotler} the out-of-band interactions can be surpressed by the use of Slepian-functions~\cite{Frey2017}.

\begin{figure}
\centering
\includegraphics[width=0.34\textwidth]{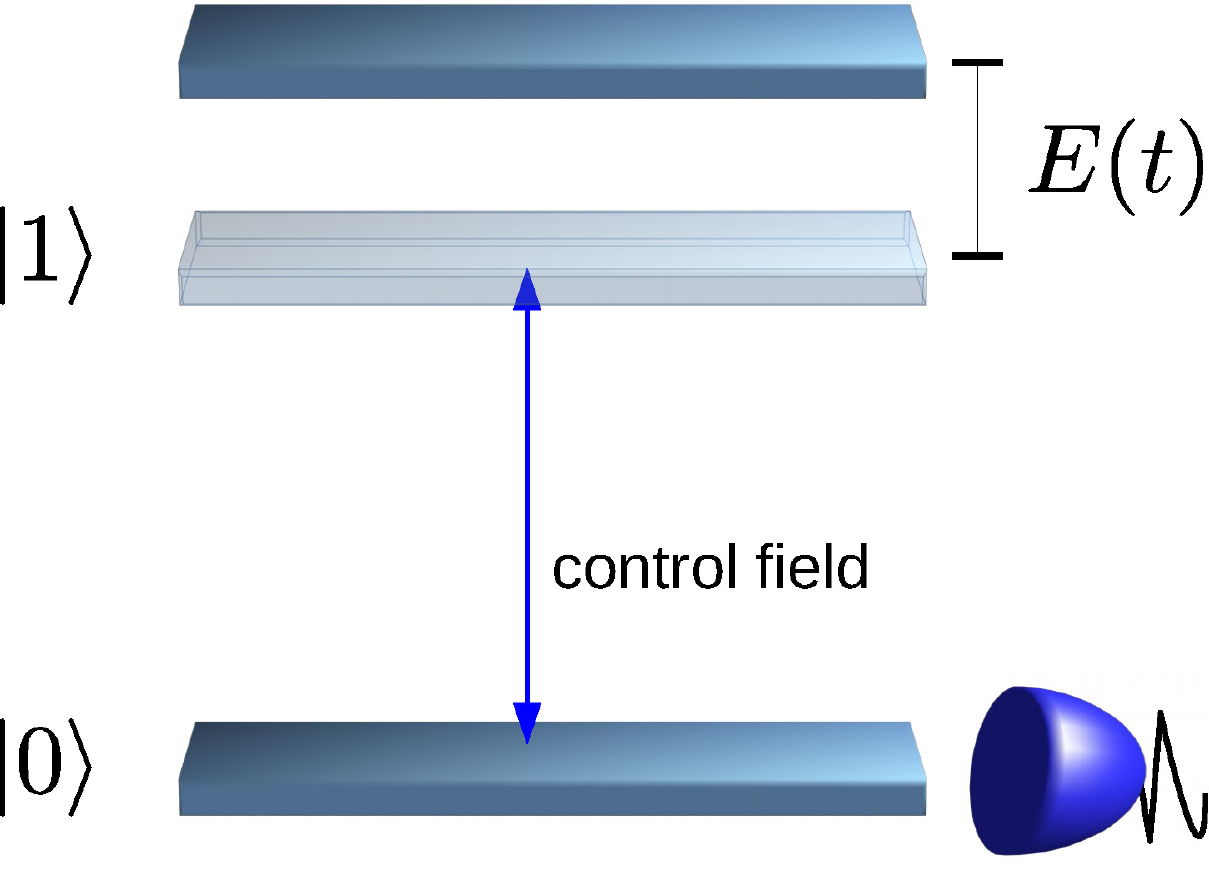}
\caption{Single qubit quantum probe. A fluctuating field $E(t)$ acts on a two-level system via $E(t)\sigma_z$. We can measure the population in the lower state $|0\rangle$ and drive the transition between the two states $|0\rangle$ and $|1\rangle$ by a control field.}\label{fig:sketch}
\end{figure}

In this paper, we introduce a reconstruction algorithm for the spectrum of a signal that is based on the orthogonalization of the applied filter functions, as well as on the use of entangled multi-qubit probes. We test the robustness of this reconstruction algorithm for two different noise models, based on dephasing and finite precision measurement, as well as on statistical noise. We quantify the latter in terms of a directional Fisher information, based on the Fisher Information Operator recently introduced in the context of stochastic quantum Zeno dynamics~\cite{MuellerFisher,MuellerSciRep} by us. Finally, we employ optimal control theory to construct filter functions that maximize this Fisher information and thus the sensitivity of the filter with respect to the signal. We show how both filter orthogonalization and filter optimization brings about a speed-up in the sensing procedure that can help to fight different forms of noise.

\section{Results}
\subsection{Filter function approach and protocols for spectrally resolved sensing}\label{sec:filter-functions}

Let us consider a two-level system, given by the two levels $|0\rangle$ and $|1\rangle$, that interacts with a fluctuating field $E(t)$ via $E(t)\sigma_z$, where $\sigma_z$ is the z-Pauli matrix. Following \cite{Degen2016,KofmanNAT2000,Alvarez2011}, we prepare the system in its lower state $|0\rangle$ and apply an initial $\pi/2$-pulse to bring it into the superposition state $(|0\rangle + |1\rangle)/\sqrt{2}$. Then, we apply a sequence of $\pi$-pulses (i.e. population flips between $|0\rangle$ and $|1\rangle$) that we describe by a pulse modulation function $y(t)\in \{-1,1\}$ that switches sign at the position of the pulses.

The system then acquires a phase $\phi(t)$ leading to the state $[\mathrm{e}^{i\phi(t)}|0\rangle + \mathrm{e}^{-i\phi(t)}|1\rangle]/\sqrt{2}$, and this phase can be measured via a population measurement after a final $\pi/2$-pulse that brings the system into the state $[(\mathrm{e}^{i\phi(T)}+\mathrm{e}^{-i\phi(T)})|0\rangle + (\mathrm{e}^{i\phi(T)}-\mathrm{e}^{-i\phi(T)})|1\rangle]/2$, where $T$ is the total operation time of the pulse sequence. The phase is determined by the fluctuating field and the modulation function and can be written as
\begin{eqnarray}
 \phi(t)=\int_0^t y(t') E(t') dt'\,.
\end{eqnarray}
If we average over many realizations of the sequence, we can describe the dephasing by a decoherence function $\chi(t)$. This decoherence function enters the probability $p(t)$ to find the system in state $|0\rangle$ upon measurement as
\begin{equation}
 p(t)=\frac{1}{2}\big(1-\mathrm{e}^{-\chi(t)}\big)
\end{equation}
and is given by
\begin{equation}
\chi(t) = \frac{1}{2}\int_0^t\int_0^t y(t')y(t'')g(t'-t'')dt'dt'',
\end{equation}
where
\begin{equation}
g(t'-t'') = \big\langle E(t')E(t'')\big\rangle
\end{equation}
is the autocorrelation function of the fluctuating field $E(t)$. The brackets $\langle \cdot\rangle$ mean averaging over the realizations of the stochastic field $E(t)$. The Fourier transform of the autocorrelation function denotes the power spectral density $S(\omega)$ of the field $E(t)$:
\begin{equation}
 S(\omega)=\int_{-\infty}^{\infty} g(\tau) \mathrm{e}^{-i\omega\tau}d\tau\,.
\end{equation}
As shown in detail in the Methods, also for continuous pulse modulations, the decoherence function $\chi(T)$ can be written in terms of $S(\omega)$ and a so-called filter function $F(\omega)$, i.e.
\begin{eqnarray}\label{eq:filterfunctions}
 \chi(T)&=\int_{0}^{\infty} S(\omega) F_T(\omega) d\omega\,.
\end{eqnarray}
In the following, we call $T$ the filter operation time. The filter function is defined as $F_T(\omega)=\frac{4}{\pi}|Y_T(\omega)|^{2}$, where
\begin{equation}
Y_T(\omega)\equiv\int_0^T y(t)e^{i\omega t}dt
\end{equation}
is the Fourier transform of the pulse modulation function. To simplify notation, we will omit the time in the subscript and identify $F\equiv F_T$, $Y\equiv Y_T$.
By engineering the pulse modulation function, we can design different filter functions $F(\omega)$ that select specific frequency ranges of the power spectral density $S(\omega)$.
In order to estimate the functional behaviour of the power spectral density $S(\omega)$, in the following we employ a set of $K$ filter functions $F_{k}(\omega)$, $k = 1,\dots,K$, each of them generated by a different sequence of $\pi$-pulses and measure the decoherence function after the application of each of these filters. We will then describe protocols that compose the functional behaviour of $S(\omega)$ from the single measurement outcomes, so that we can reconstruct $S(\omega)$ in a range $\omega\in[0,\omega_c]$ for some cut-off frequency $\omega_c$.

We first introduce a novel protocol, which is based on the orthogonalization of the filter functions $F_{k}(\omega)$, $k = 1,\dots,K$. For the sake of brevity, from here on we will denote this new protocol with the acronym Filter Orthogonalization (FO) protocol. Then, we recall the Alvàrez-Suter (AS) protocol~\cite{Alvarez2011} that we will use as a benchmark to test the performance of the new approach. Finally, we will introduce a fidelity measure to evaluate the performance of the protocols.

\subsubsection{Filter Orthogonalization protocol}

The filter functions of the FO protocol in principle can be chosen arbitrarily. Here we choose to base them on equidistant pulse sequences and design the latter by positioning the $\pi$-pulses at the zeros of $\cos[\omega_{max}\frac{k-1}{K}t']$, with $k = 1,\dots,K$.  Note that this includes also a sequence without any $\pi$-pulse (for $k=1$) corresponding to a filter function sensitive to the frequency $\omega = 0$. Instead, $\omega_{max}$ is a constant that determines the bandwidth within which we can analyse the spectrum of the signal (the field $E(t)$), and the filter corresponding to the highest frequency is centered around $(1-1/K)\omega_{max}$. The population measurements at the end of the filter application will yield the coefficients
\begin{equation}
c_k = \int_0^\infty S(\omega) F_k(\omega)d\omega\,,
\end{equation}
which is the value of the decoherence function $\chi(T)$ at the end of the application of the filter $F_k$. We then calculate the $K\times K$ matrix $A$, whose matrix elements
\begin{eqnarray}
A_{kl} = \int_0^{\omega_c} F_k(\omega)F_l(\omega) d\omega
\end{eqnarray}
quantify the overlap in the frequency domain between the filter functions $F_{k}(\omega)$ and $F_{l}(\omega)$ ($k,l\, \in\, 1,\dots, K$). We truncate the integral at the cut-off frequency $\omega_c$, since we want to analyse $S(\omega)$ only in the interval $[0,\omega_c]$. The matrix $A$ is symmetric and can be orthogonalized by means of the following transformation:
\begin{equation}
V A V^{\mathrm{T}}=  \Lambda\,,
\end{equation}
where $\Lambda\equiv\mathrm{diag}(\lambda_1,\dots,\lambda_K)$ are the eigenvalues of $A$ and $V$ is an orthogonal matrix.
The $K$ filters span a $K$-dimensional function space that has an orthonormal basis
\begin{eqnarray}
 \tilde{F}_k(\omega)=\frac{1}{\sqrt{\lambda_k}}\sum_{l=1}^K V_{kl}F_l(\omega),\quad k=1,\dots, K\\
 \int_0^{\omega_c} \tilde{F}_k(\omega) \tilde{F}_l(\omega)=\delta_{kl}\,,
\end{eqnarray}
with the Kronecker-Delta $\delta_{kl}$.
If we expand the spectral density $S(\omega)$ in this orthogonal basis, we obtain the coefficients
\begin{eqnarray}
\widetilde{c}_{k} = \int_0^\infty S(\omega) \widetilde{F}_k(\omega)=\frac{1}{\sqrt{\lambda_k}}\sum_{l=1}^K V_{kl}c_l\,.
\end{eqnarray}
Hence, we obtain an estimate $\widetilde{S}(\omega)$ of the power spectral density $S(\omega)$ given by the expansion
\begin{eqnarray}\label{eq:FO-expansion}
\widetilde{S}(\omega)=\sum_{k=1}^{K}\widetilde{c}_{k}\widetilde{F}_{k}\,.
\end{eqnarray}

\subsubsection{Alvàrez-Suter protocol}

For the AS protocol we use as well $K$ equidistant pulse sequences. However, due to the requirements of this protocol, this time we position the $\pi$-pulses at the zeros of
\begin{eqnarray}
 \sin\big[\omega_{max}\frac{k}{K}t'\big],\,\,k=1,\dots, K\,.
\end{eqnarray}
As a consequence, the minimal frequency that can be resolved is $\omega_{max}/K$, while the maximual frequency that can be resolved is exactly $\omega_{max}$. Indeed, the protocol will yield the value of $S(\omega)$ at the $K$ discrete points $\omega_k=\omega_{max}\frac{k}{K}$.

To reconstruct $S(\omega)$ pointwise, we then employ the AS protocol~\cite{Alvarez2011}, where long pulse sequences (ca. 30-100 pulses) produce a Dirac-delta-like shape of the filter functions. The FO protocol, instead, does not rely on this Dirac-delta-shape filters and thus can in principle work with a smaller number of pulses (smaller filter operation time) and reconstructs the power spectral density $S(\omega)$ as a continuous function. In the following, unless specified otherwise, for both protocols we choose $K=20$ and $\omega_{c}=10$. For the FO protocol, we choose $\omega_{max}=11.5$ to avoid border effects near to the cut-off frequency. For the AS protocol, instead, we choose $\omega_{max}=\omega_c=10$ so that the $K$ values of $S(\omega)$ obtained by this protocol cover the desired bandwidth.

\subsubsection{Estimation Fidelity}

To compare the two estimation protocols and different sets of parameters, we introduce a fidelity measure. This is not straightforward, since the AS protocol reconstructs $S(\omega)$ only in a set of discrete points $\omega_{k}=\omega_{max}\frac{k}{K}$, $k = 1,\dots K$, while the FO protocol reconstructs a continuous function.
We choose to compare the values of $S(\omega)$ only on the mentioned set of discrete points, and define the fidelity as
\begin{eqnarray}
 \mathcal{F}=\frac{1}{K}\sum_{k=1}^K \frac{S(\omega_k)\tilde{S}(\omega_k)}{\Vert S(\omega)\Vert \Vert\tilde{S}(\omega)\Vert},
\end{eqnarray}
where $\tilde{S}(\omega)$ is the reconstructed power spectral density and the norm $\Vert S(\omega)\Vert$ is given by
$$\Vert S(\omega)\Vert^2=\frac{1}{K}\sum_{k=1}^K S(\omega_k)^2$$
for the power spectral density and analogously for its estimate $\tilde{S}(\omega)$.

\subsection{Robust noise sensing under measurement noise and dephasing}

The estimation of the power spectral density $S(\omega)$ relies on the measurement of the probability $p_{k} = \frac{1}{2}(1-e^{-c_k})$, which quantifies the overlap between $S(\omega)$ and each filter function $F_{k}(\omega)$, $k = 1,\ldots,N$, given by the coefficients $c_{k}=\int S(\omega)F_{k}(\omega)d\omega$. The values of the $p_k$ are measured via state population measurements, which are unavoidably affected by external sources of error, due to the presence of imperfections of the measurement device, of statistical errors and of detector noise contributions.

In this section, we will model two essential sources of error in the state population measurement. In particular, we assume that each measurement of $p_{k}$ is affected by an absolute error $\Delta p_k$, which is independent from the value of $p_{k}$ (and thus its statistics is independent from $k$). This error $\Delta p_k$ is chosen randomly from a uniform distribution defined by the interval $\Delta p_k\in [-\Delta p_{max},\Delta p_{max}]$, where $\Delta p_{max}\geq 0$ is the maximal absolute value admissible by $\Delta p_k$. Furthermore, we assume also that the relative phase between the two probe qubit states ($|0\rangle$ and $|1\rangle$, respectively) is additionally damped by a dephasing contribution (different from the influence of the field $E(t)$). This additional dephasing models the instability/natural imperfection of the probe system and is described by the rate $\Gamma$, such that for a filter operation time $T$ the decoherence function increases by $\Gamma T$ with respect to the case without dephasing. We thus find for the population measurement
\begin{eqnarray}\label{eq:p-measured}
p_{k} = \frac{1}{2}\left(1-\mathrm{e}^{-c_{k}-\Gamma T}\right)\,,
\end{eqnarray}
and, as a consequence, for a small error $\Delta p_k$ it holds that
\begin{eqnarray}\label{delta_p}
\Delta p_k = \mathrm{e}^{-c_k-\Gamma T}\Delta c_k\,,
\end{eqnarray}
which has been obtained by performing the derivative of Eq.\eqref{eq:p-measured}. In Eq.\eqref{delta_p}, $\Delta c_{k}$ is the resulting error in the coefficient $c_{k}$. Since we want to determine the functional shape of $S(\omega)$, the relevant error is the relative error $\Delta c_{k}/c_{k}$ (the absolute value of the coefficients $c_{k}$ can usually be tuned by changing the distance of the probe to the source of the signal $S(\omega)$, which results in a change in the intensity of the field $E(t)$ at the site of the probe). For this relative error we find
\begin{eqnarray}
\frac{\Delta c_{k}}{c_{k}}=\frac{\mathrm{e}^{c_{k}}}{c_{k}}\mathrm{e}^{\Gamma T}\Delta p_k\,,
\end{eqnarray}
which takes its minimum for $c_{k} = 1$, with a quite flat behaviour (about $10\,\%$ increase) between about $0.5$ and $1.5$. Accordingly, if possible, the measurement has to be performed in this regime of maximum sensitivity and we will do this in the simulations. Let us observe, moreover, that the relative error scales exponentially with respect to the filter operation time $T$ and the dephasing rate $\Gamma$.

\begin{figure}
 \includegraphics[scale = 1]{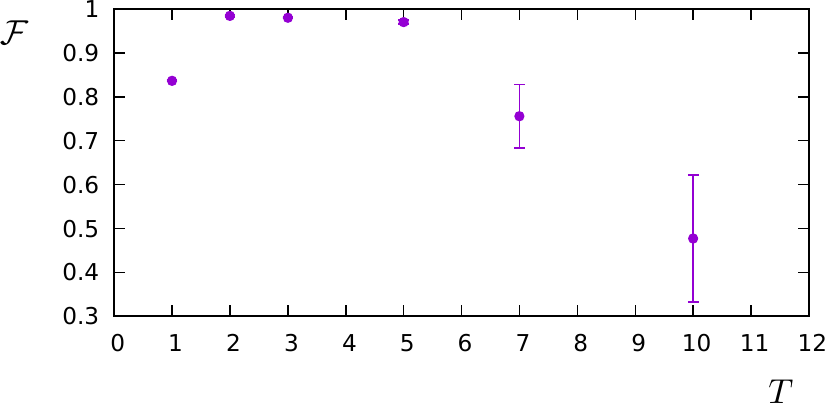}
 \caption{The figure shows how the fidelity of the estimation of $S(\omega)$ scales with the filter operation time $T$. The simulations were performed by using the FO protocol, with dephasing $\Gamma = 0.4$, and detector noise $\Delta p_{max} = 0.01$.}
\label{fig:fidelity-vs-time}
\end{figure}
\begin{figure}
 \begin{overpic}[scale=1]{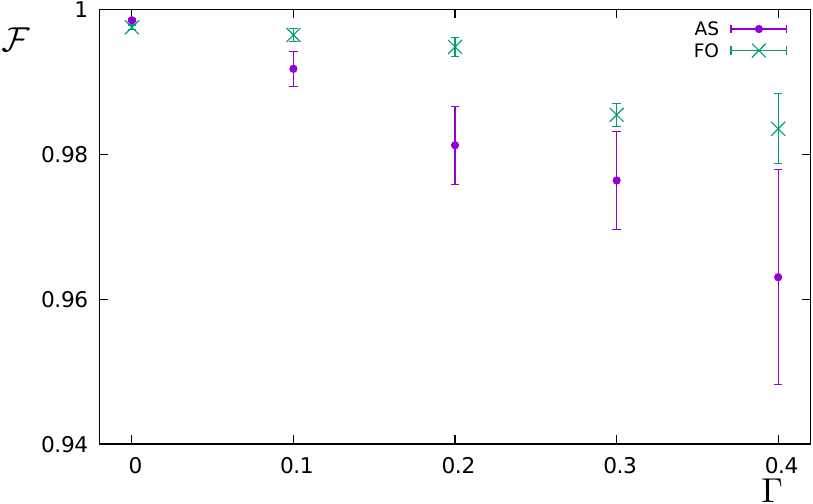}
\end{overpic}
 \caption{The figure shows the fidelity for the estimation of the power spectral density $S(\omega)$ as a function of the dephasing rate $\Gamma$. For each value of $\Gamma$, we choose the filter operation time such that the fidelity is maximized. This optimal time is obtained from a scan of different values of the filter operation time $t$ both for the AS and FO protocol as in the example of Fig.~\ref{fig:fidelity-vs-time}. The optimal filter operation time depends on $\Gamma$ and on the protocol, and was found to be in the range $[5,25]$ for the AS protocol and $[2,5]$ for the FO protocol.}
 \label{fig:fidelity-vs-dephasing}
\end{figure}
We have studied the performance of the FO protocol in comparison with the AS protocol by simulating the measurement of the power spectral density $S(\omega)=\frac{S_0}{1+(\omega-2)^2}+\frac{0.7S_0}{1+2(\omega-6)^2}$, which is a spectrum that can arise for example from the spontaneous decay at two different frequencies. For the error model, we have assumed a measurement noise given by $\Delta p_{max} = 0.01$ and we have chosen $S_0\propto T$ in such a way that the probe operates in the maximum sensitivity range ($c_{k}\approx 1$) for all choices of $T$ and most of the $K$ filters (we consider the same $S_0$ and $T$ for all $K$ filters, but not all of them have the same overlap with $S(\omega)$). Let us observe that for a given dephasing rate $\Gamma$ the time-exponential increase of the measurement error $\Delta c_{k}/c_{k}$ will favour smaller values of the filter operation time $T$. On the other side, if we choose $T$ too small, then the filters lose their frequency selectivity as they flatten out in the frequency domain. Fig.~\ref{fig:fidelity-vs-time} shows the scaling of the fidelity with the filter operation time $T$ for the FO protocol for a dephasing rate $\Gamma = 0.4$. For each value of $\Gamma$ and for both sensing protocols, we have prepared such a graph to determine the optimal value of $T$ out of the tested values, e.g. $T=2$ out of the tested values $T=1,2,3,5,7,10$ in the example of Fig.~\ref{fig:fidelity-vs-time}.
\begin{figure}[h]
 \begin{overpic}[scale=0.85]{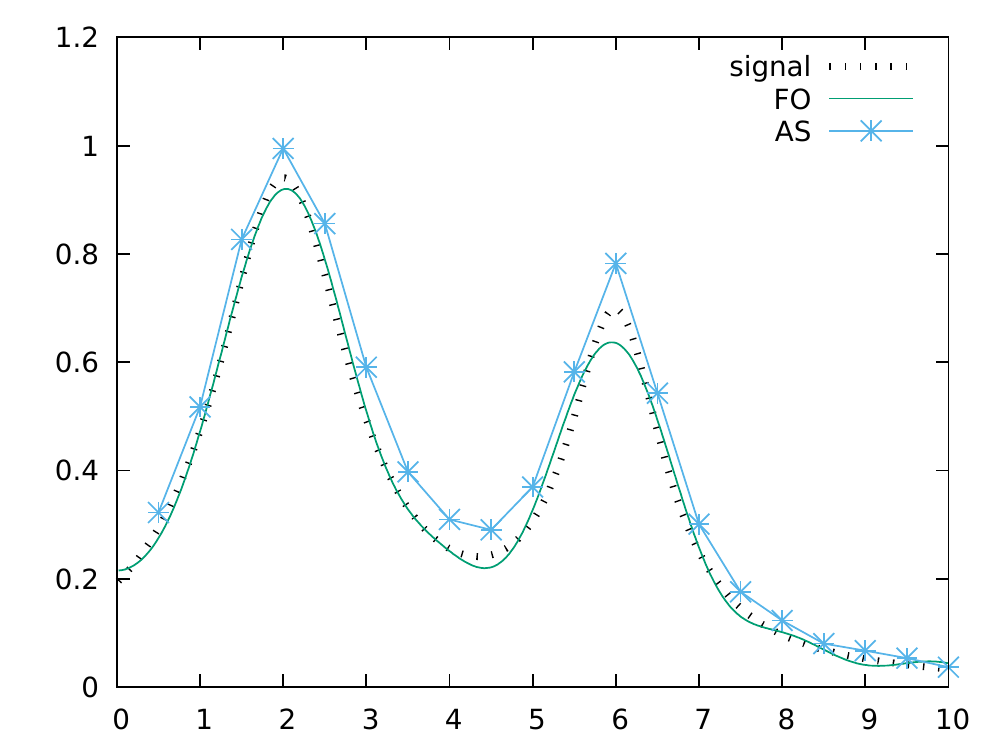}
 \put(-3,65){$S(\omega)$}
 \put(80,-2){$\omega$}
 \end{overpic}
 \caption{Estimation of the power spectral density $S(\omega)$ by applying the AS and the FO protocols, where the measurements are subjected to a detector noise of $\Delta p_{max}=0.01$ (in the absence of additional dephasing, i.e. $\Gamma=0$).}
 \label{fig:dephasing0}
\end{figure}
\begin{figure}[h]
 \begin{overpic}[scale=0.85]{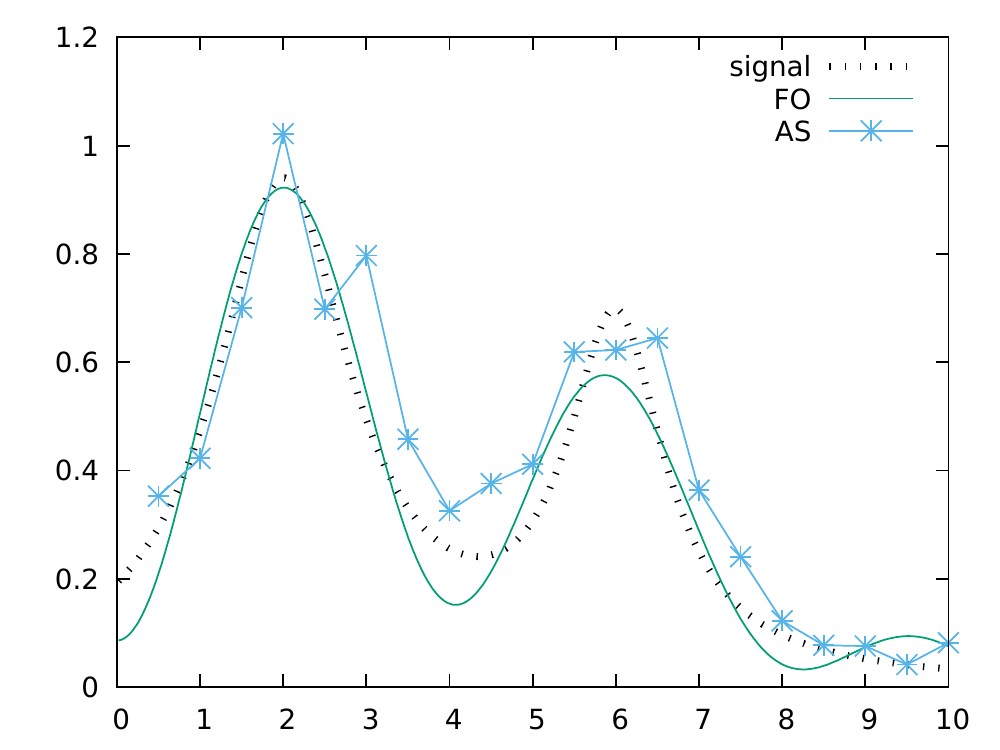}
 \put(-3,65){$S(\omega)$}
 \put(80,-2){$\omega$}
 \end{overpic}
 \caption{Estimation of the power spectral density $S(\omega)$ by applying the AS and FO protocols, where (as in Fig.~\ref{fig:dephasing0}) the measurements are subjected to a detector noise of $\Delta p_{max}=0.01$. Additionally, the probes are affected by a dephasing given by the dephasing rate $\Gamma=0.4$.}
 \label{fig:dephasing4}
\end{figure}
We reconstruct the spectral density $S(\omega)$ for this optimal choice of $T$ and show the fidelity of this estimate as a function of the dephasing $\Gamma$ in Fig.~\ref{fig:fidelity-vs-dephasing}.  Finally, the reconstructed functional shape of the power spectral density $S(\omega)$ obtained by each of the sensing protocols is shown in Fig.~\ref{fig:dephasing0} for $\Gamma = 0$ and in Fig.~\ref{fig:dephasing4} for $\Gamma=0.4$, respectively. Note that if the measurements are affected by noise, for the FO protocol we can improve the estimate by omitting the terms with the smallest eigenvalues $\lambda_k$ of the matrix $A$ in the expansion of Eq.~\eqref{eq:FO-expansion}. In the simulations, depending on $T$ we have used between 10-20 terms out of $K=20$ (the larger $T$, the less eigenvalues are omitted) in order to minimize the effect of the noise.

\subsubsection{Application: NV-centers in diamond}

The above examples are calculated with dimensionless variables and the results hold with suitable scaling of the parameters. As an example application, we can consider a nitrogen-vacancy defect in diamond. Typical dephasing rates are of the order of $1/\Gamma=100\,\mathrm{\mu s}$ \cite{Bar-Gill}. By fixing the dephasing rate, the units of all other parameters are defined, so that we can now interpret Fig.~\ref{fig:dephasing4} as an example with NV-centers in diamond with parameters $1/\Gamma=100\,\mathrm{\mu s}$, $\Delta p_{max}=0.01$, and $\omega\in [0,2\pi\times 0.4\,\mathrm{kHz}]$. By taking such values for the noise parameters, the corresponding operation time is, respectively, equal to $T=80\,\mathrm{\mu s}$ for the FO protocol and $T=200\,\mathrm{\mu s}$ for the AS protocol.

\subsection{Fighting spectral leakage via entangled multi-qubit probes}

In this section, we introduce estimation protocols employing entangled multi-qubit probes and show how they can be used to fight spectral leakage in the sensing protocol. Spectral leakage is a known source of error that arises from the fact that the support of the filter functions is larger than the bandwidth they can analyse~\cite{Alvarez2011,Norris2016}. As a consequence, spectral contributions of the signal $S(\omega)$ outside the analysed bandwidth, but within the support of the filters, can disturb the reconstruction of the signal also within the analysed bandwidth.

The key idea of our entangled multi-qubit probes is that instead of applying the initial $\pi/2$ pulses, we prepare an $N$ qubit system in the GHZ state
\begin{equation}
\frac{|0...0\rangle + |1...1\rangle}{\sqrt{2}}.
\end{equation}
Then, during the main part of the protocol, we apply single qubit pulses to each qubit individually. The pulse modulation function $y(t)$ is then generalized to the single qubit modulation function $y_i(t)$ ($i=1,\dots, N$), where a switch of sign of $y_j(t)$ indicates the application of a single qubit $\pi$-pulse on qubit $j$. The final filter function for $N$ qubits then reads
\begin{equation}
 F(\omega)=\frac{4}{\pi}|Y(\omega)|^2=\frac{4}{\pi}\bigg|\int_0^T \sum_{j=1}^N y_j(t) e^{i\omega t}dt\bigg|^2\,.\\
\end{equation}
We can then use these multi-qubit probe filters for the FO protocol. For more details we refer to the Methods.

\begin{figure}
 \includegraphics[width=0.47\textwidth]{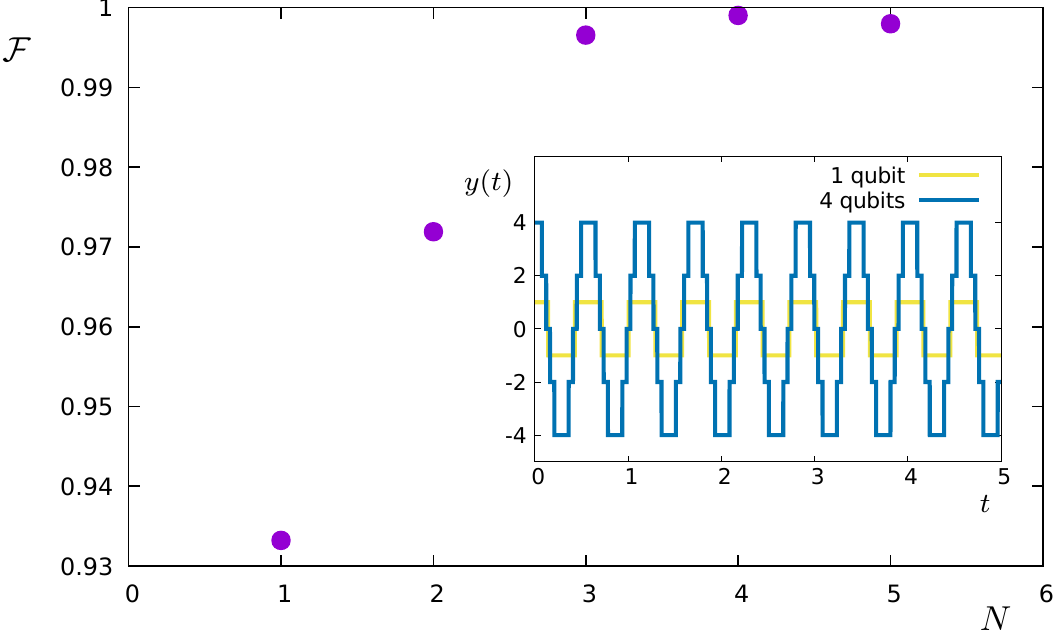}
 \caption{Fidelity vs number of qubits in the multi-qubit probe. The inset shows the pulse modulation function $y(t)=\sum_{j=1}^N y_j(t)$ for $N=1$ qubit as well as for $N=4$ qubits for one specific choice of filter frequency.}
 \label{fig:fidelity-vs-nqubit}
\end{figure}
\begin{figure}
\begin{overpic}[scale=.85]{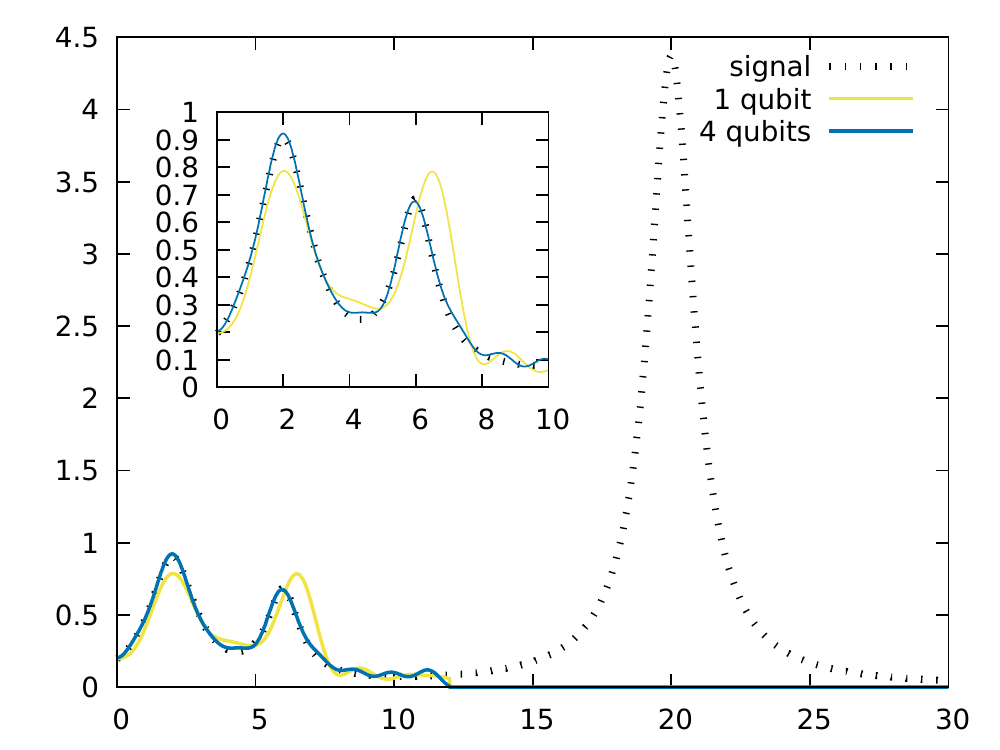}
 \put(-3,65){$S(\omega)$}
 \put(80,-2){$\omega$}
\end{overpic}
 \caption{Spectrum estimate obtained by a single qubit probe compared to the one obtained using a four-qubit probe. The original spectrum has a significant contribution outside the filter range and thus causes an error in the estimate obtained by the single qubit. The inset is a zoom into the analysed range ($\omega\in[0,10]$). The target spectrum is given by $S(\omega)=\frac{S_0}{1+(\omega-2)^2}+\frac{0.7S_0}{1+2(\omega-6)^2}+\frac{5S_0}{1+(\omega-20)^2}$.}
 \label{fig:spectrum-vs-nqubit}
\end{figure}
We test the multi-qubit filters by reconstructing $$S(\omega)=\frac{S_0}{1+(\omega-2)^2}+\frac{0.7S_0}{1+2(\omega-6)^2}+\frac{5S_0}{1+(\omega-20)^2}$$
in a range $\omega\in [0,\omega_c=10]$ with filters designed for that range. The first two terms are as in the previous section. The last term has its main contribution outside the analysed range and thus potentially leads to spectral leakage. We choose its amplitude to be large enough to effectively disturb the measurement of the spectrum in the range $\omega\in [0,\omega_c=10]$ using the methods presented in the previous section.

Fig.~\ref{fig:fidelity-vs-nqubit} shows how the estimation fidelity (FO protocol) scales with the number of qubits in the multi-qubit probe. Note that this is done only for the FO protocol since the AS protocol does not allow for a straightforward extension to multiple qubits as it relies on Dirac-Delta-shaped filters. Fig.~\ref{fig:spectrum-vs-nqubit} gives an insight to the dependency of the estimation fidelity on the number of qubits: The peak of the power spectral density outside the filter range causes a distortion of the reconstruction when using only one qubit (spectral leakage). The reason is that the pulse modulation function introduces also higher frequencies (i.e. outside the analysed bandwidth or filter range) due to the stepwise modulation. A multi-qubit probe allows for intermediate steps and thus suppresses these high-frequency contributions in the filters. As a consequence, the peak outside the filter range does not significantly disturb the reconstruction within the analysed range. This is true already for only 3-5 qubits.

\subsubsection{Application: Trapped Ions}
\begin{figure}
 \includegraphics[width=0.47\textwidth]{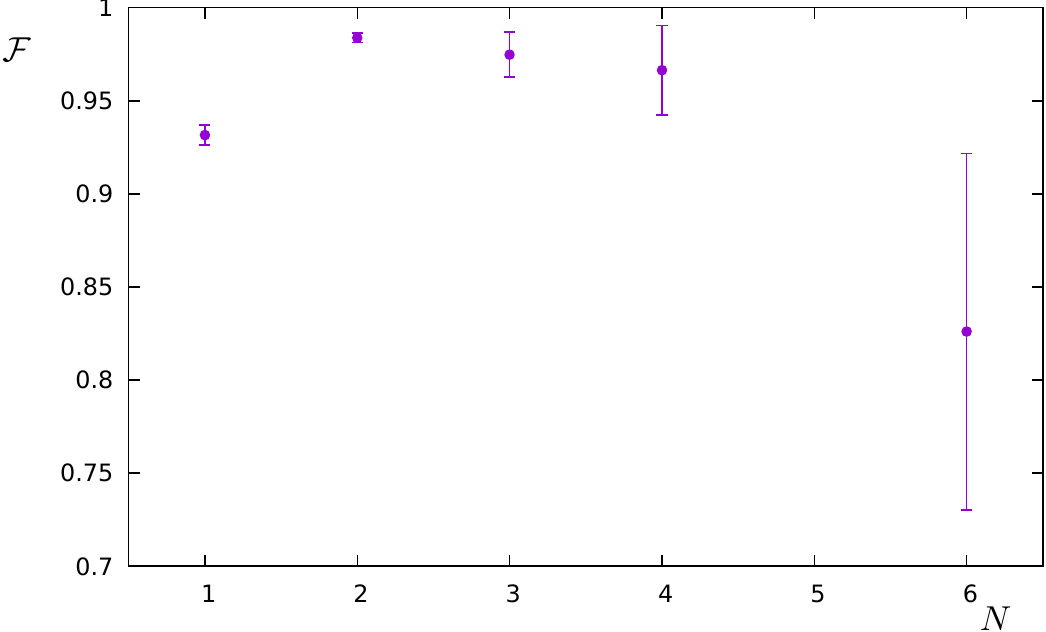}
 \caption{Fidelity vs number of qubits in the multi-qubit probe considering state preparation and dephasing errors corresponding to the chain of trapped ions that is discussed in the text.}
 \label{fig:fidelity-vs-nqubit-ion}
\end{figure}
The essential ingredient in this section is the creation of the GHZ-states. These states can be conveniently generated in linear ion traps via a M\o{}lmer-S\o{}rensen-gate operation \cite{Monz2011}. Each ion represents one qubit and the gate operation generates a GHZ-state on the spin states of the ions. While the gate is very robust with respect to the state of the vibrational mode, still the states are not perfectly generated, especially for more than two qubits. In our model, we encode this state-preparation error in the value of $\Delta p_{max}$. Inspired by the experimental values~\cite{Monz2011} we choose $\Delta p_{max}=0.01, 0.02, 0.03, 0.04, 0.1$ for $N=1,2,3,4,6$ qubits (note that we omit $N=5$, since in the reference no value for the fidelity for $N=5$ is presented), respectively and a dephasing rate $\Gamma=0.01\,\mathrm{ms}^{-1}$. We simulate the measurement of the signal $$S(\omega)=\frac{S_0}{1+(\omega-\omega_1)^2}+\frac{0.7S_0}{1+2(\omega-\omega_2)^2}+\frac{5S_0}{1+(\omega-\omega_3)^2}$$
in a range $\omega\in [0,\omega_c]$ with filters designed for that range.
The frequency parameters are $\omega_1=2\pi\times 1\,\mathrm{kHz}$,
$\omega_2=2\pi\times 3\,\mathrm{kHz}$,
$\omega_3=2\pi\times 10\,\mathrm{kHz}$, and $\omega_c=2\pi\times 5\,\mathrm{kHz}$.
Fig.~\ref{fig:fidelity-vs-nqubit-ion} shows the resulting fidelity of reconstruction. The optimal filter operation time turned out to be $T=10\,$ms for $N=1$ qubit and $T=4\,$ms for $N\geq 2$ qubits. Smaller filter operation times do not produce the frequency range accurately enough. Larger filter operation times are affected by dephasing, especially collective dephasing when more than one qubit is used. The highest fidelity is achieved for only two qubits with $\mathcal{F}=0.984\pm0.0025$. This value could be improved by correcting for deterministic errors in the state preparation, reducing the state preparation error, or reducing the dephasing rate. By doing so, the optimum could be reached for $N>2$.

\subsection{Optimal control filters -- via Fisher information}

By measuring the survival probability at the end of the filter sequence, we can indeed determine the decoherence function $\chi(t)$ and the following functional
\begin{equation}
\xi=\int_0^{\infty} S(\omega) F(\omega) d\omega/\Vert F(\omega)\Vert_c\,,
\end{equation}
where we introduce the continuous norm of the filter function $\Vert F(\omega)\Vert_c$, and the power spectral density $\Vert S(\omega)\Vert_c$ by
\begin{eqnarray}
 \Vert F\Vert_c^2= \int_0^{\omega_c} |F(\omega)|^2 d\omega\nonumber\\
 \Vert S\Vert_c^2= \int_0^{\omega_c} |S(\omega)|^2 d\omega\,.
\end{eqnarray}
Note, that $\xi=\chi(t)/\Vert F\Vert_c$ and that we know $F(\omega)$, but not $S(\omega)$.
If we now optimize the modulation function $y(t)=\sum_{j=1}^N y_j(t)$ in order to maximize $\xi$, we will change $\frac{F(\omega)}{\Vert F\Vert_c}$ until it points in the same direction in function space as $S(\omega)$. I.e., at the maximum point we will find
\begin{equation}
 \frac{F(\omega)}{\Vert F\Vert_c}\approx \frac{S(\omega)}{\Vert S\Vert_c}\,.
\end{equation}
Note that we can choose the cut-off frequency $\omega_c$ in such a way that the support of the signal is not truncated, $\mathrm{supp}\, S(\omega)\subset [0,\omega_c]$. Alternatively, we can add an additional term to the control objective $\xi$ imposing a constraint on the optimized filters such that they have a vanishing contribution for $\omega>\omega_c$.

As shown in the Methods section, this optimization procedure corresponds to the maximization of the Fisher Information associated with the sensitivity of the measurement in the direction (of the functional space \cite{MuellerFisher}) $S(\omega)$, and thus to the maximum sensitivity of the filter with respect to the signal. A similar approach for single-parameter estimation has allowed to determine the characteristic width of the spectrum with maximum sensitivity \cite{Zwick2016}.

\begin{figure}
\centering
\includegraphics[width=0.47\textwidth]{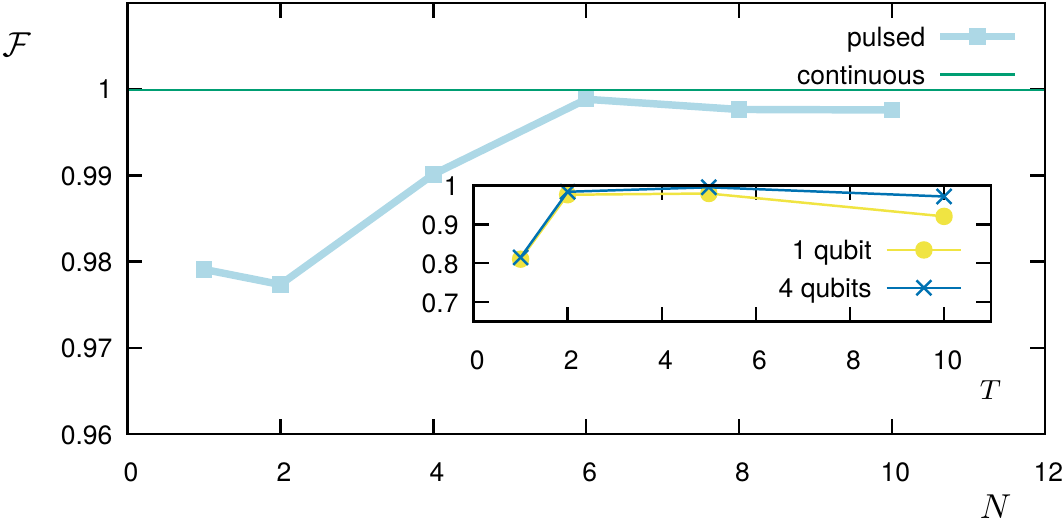}
 \caption{Optimal fidelity $\mathcal{F}$ vs. number of qubits $N$, when the optimal-control filter reconstructs the power spectral density for $T=5$ (light blue). The green line shows the near-unity fidelity reached with continuous pulse modulations (an infinite number of qubits or a continuous modulation of $\Omega(t)$). The inset shows the fidelity obtained with only 1 qubit (violet) and with 4 qubits (blue) as a function of $T$.}\label{fig:fidelity-pulse-shaping}
\end{figure}

We test the optimization numerically by trying to design a filter that maximizes the functional $\xi$ for two examples of a given spectral density $S(\omega)$: we fix $S(\omega)$ and perform the optimization of the control pulses via the DCRAB algorithm \cite{Rach} that straightforwardly allows to include constraints on the pulse bandwidth (or $\omega_c$) as well as the additional constraint reflecting the use of $\pi$-pulses for single or multi-qubit probes.
Fig.~\ref{fig:fidelity-pulse-shaping} shows the fidelity $\xi/\Vert S\Vert_c$ for an example of a Lorentzian power spectral density
$$S(\omega)=\frac{S_0}{1+(\omega-2)^2}$$
as a function of the number $N$ of qubits employed for the filtering. The green horizontal line shows the near-unity fidelity obtained in the ideal limit of a continuous pulse modulation (an infinite number of qubits or a continuous modulation of $\Omega(t)$ -- see Methods). The inset shows how the fidelity scales with the pulse duration $T$ for 1 qubit (yellow) and 4 qubits (blue). Note that in both cases the fidelity peaks at $T=5$. We choose this optimal value for the time $T$ also for all other optimal-control filter (OCF) results presented hereafter. However, in general this optimal value of $T$ depends on the desired frequency resolution: if the signal is less smooth (as a function of $\omega$) than in our example, the optimal value of $T$ will be larger; if it is smoother, the optimal $T$ will be smaller.

\begin{figure}
\centering
 \includegraphics[width=0.47\textwidth]{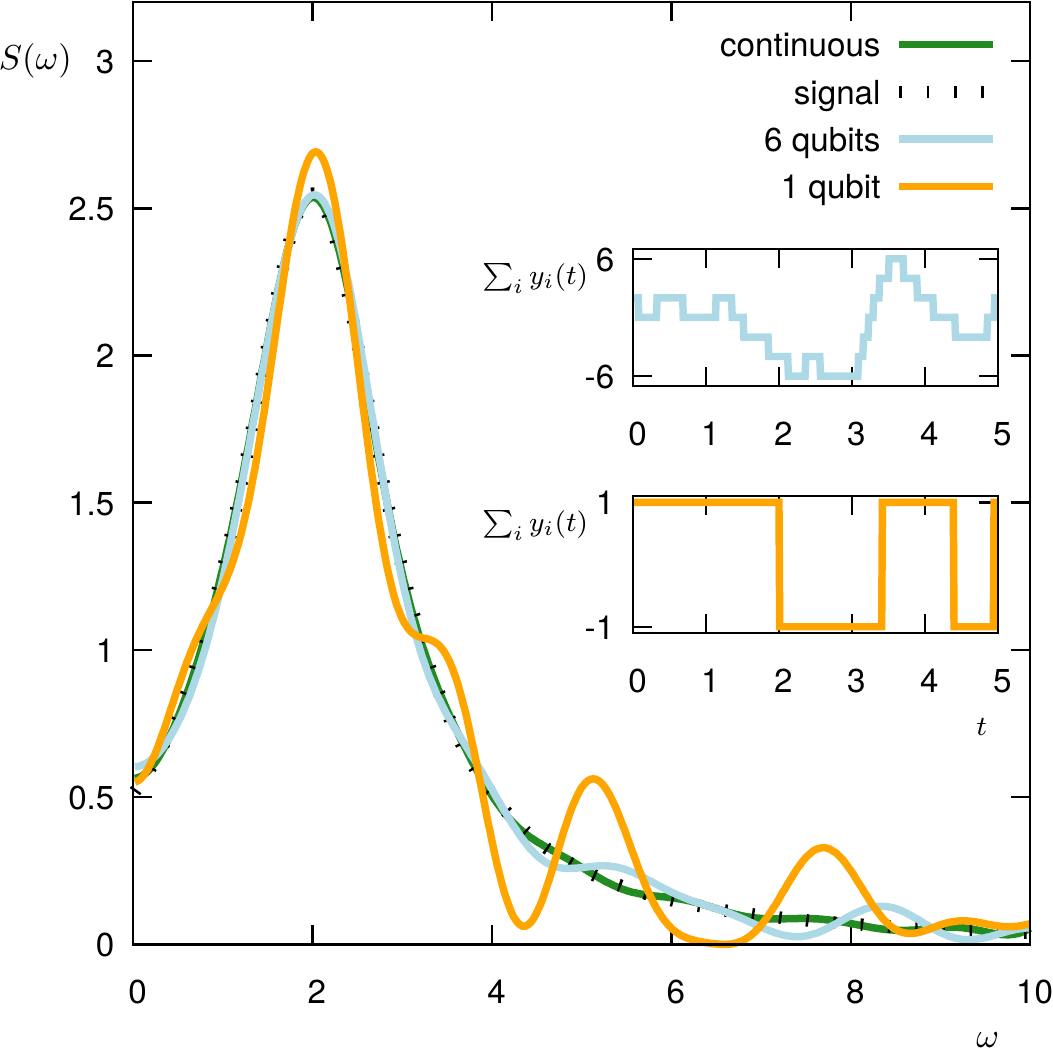}
 \caption{The (normalized) power spectral density $S(\omega)=\frac{S_0}{1+(\omega-2)^2}$ (black dashed) as reconstructed by the optimal-control filter using 6 qubits (light blue solid line) and using just 1 qubit (orange solid line). The green line is the reconstruction obtained by a continuous pulse modulation. The insets show the corresponding modulation functions for the two discrete cases (top: 6 qubits; bottom: 1 qubit).}\label{fig:pulse-shape}
\end{figure}

Fig.~\ref{fig:pulse-shape} shows the corresponding approximations of $S(\omega)$ obtained by $1$ (yellow) and $6$ (blue) qubits, respectively. The green line shows the result in the limit of a continuous pulse modulation. The insets show the shape of the modulation functions for the two discrete cases.

\begin{figure}
\centering
 \includegraphics[width=0.47\textwidth]{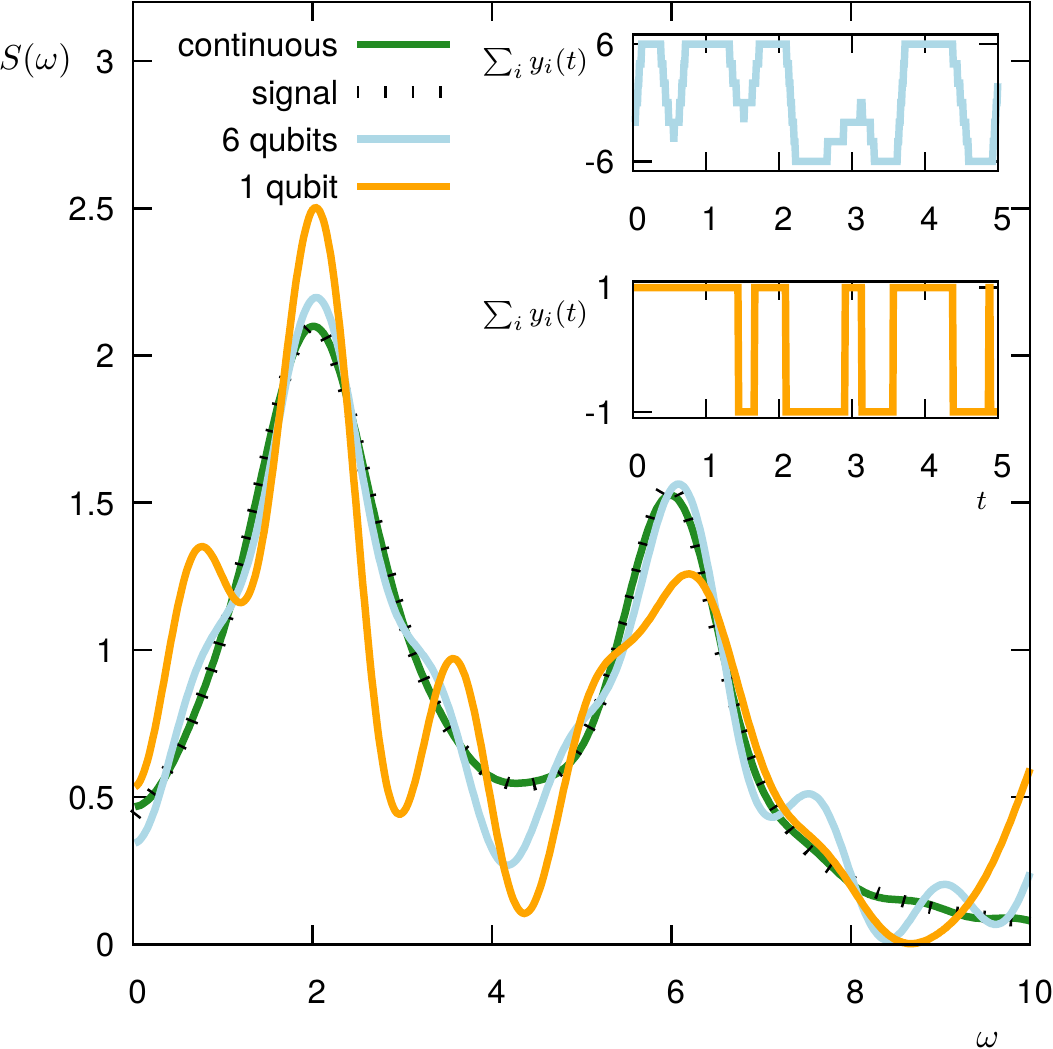}
 \caption{The (normalized) power spectral density $S(\omega)=\frac{S_0}{1+(\omega-2)^2}+\frac{0.7S_0}{1+2(\omega-6)^2}$ (black dashed) as reconstructed by the optimal-control filter using 6 qubits (light blue solid line) and using just 1 qubit (orange solid line). The green line is the reconstruction obtained by a continuous pulse modulation. The insets show the corresponding modulation functions for the two discrete cases (top: 6 qubits; bottom: 1 qubit).}\label{fig:pulse-shape-double}
\end{figure}

Fig.~\ref{fig:pulse-shape-double} shows corresponding the results for an example of a double Lorentzian power spectral density
$$S(\omega)=\frac{S_0}{1+(\omega-2)^2}+\frac{0.7S_0}{1+2(\omega-6)^2},$$
with very similar conclusions.

\subsection{Fast detection of time-dependent spectra}

\begin{figure}
 \centering
 \includegraphics[width=0.5\textwidth]{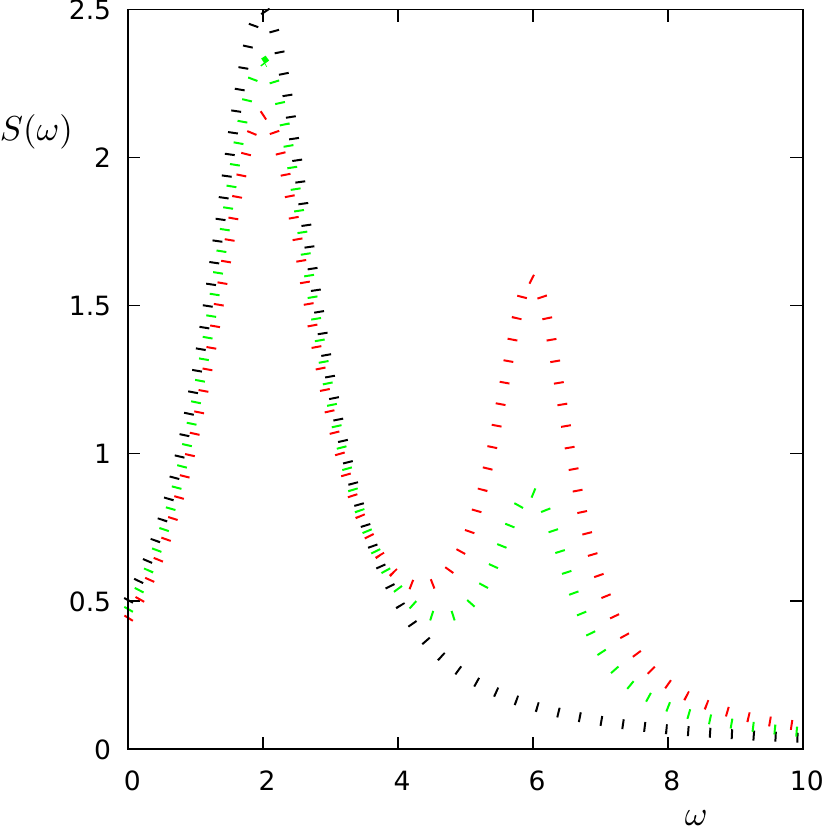}
 \caption{Composite signal. The figure shows the composite signal $S(\omega,t)=s_1(t) S_1(\omega) + s_2(t) S_2(\omega)$ for $s_1(t)=1$, $s_2(t)=0$ (black), $s_1(t)=0$, $s_2(t)=1$ (red), and $s_1(t)=0.5$, $s_2(t)=0.5$ (green). By measuring $S(\omega,t)$ with adequate protocols we want to determine $s_1(t)$ and $s_2(t)$ as a function of time.}\label{fig:composite-signal}
\end{figure}

We now consider a complex dynamical system that can exhibit two different competing processes, each leading to a different spectral behaviour of an external field (signal). We further assume that the system is not stable but rather oscillating between the two regimes and model the composite signal as
\begin{equation}\label{eq:ansatz-composite-signal}
 S(\omega,t)=s_1(t) S_1(\omega) + s_2(t) S_2(\omega)\,,
\end{equation}
where
\begin{eqnarray*}
s_1(t)=\sin^2(\omega_{osc} t),\\
s_2(t)=\cos^2(\omega_{osc} t),\\
S_1(\omega)=\frac{S_{1,0}}{1+(\omega-2)^2},\\
S_2(\omega)=\frac{S_{2,0}}{1+(\omega-2)^2}+\frac{0.7S_{2,0}}{1+2(\omega-6)^2}\,.
\end{eqnarray*}
Here, $S_1(\omega)$ and $S_2(\omega)$ are the signals corresponding to the two regimes and $S_{1,0}$, $S_{2,0}$ are normalization constants. The time-dependent coefficients $s_1(t)$ and $s_2(t)$ model the oscillation between the two spectral components. Fig.~\ref{fig:composite-signal} shows the signal $S(\omega,t)$ for different time instances corresponding to the value of the coefficients $s_1(t)=1$, $s_2(t)=0$ (black), $s_1(t)=0$, $s_2(t)=1$ (red), and $s_1(t)=0.5$, $s_2(t)=0.5$ (green).

In this section, we want to study how these time-dependent coefficients can be measured by the methods developed in this paper, under the assumption that we know the two single components $S_1(\omega)$ and $S_2(\omega)$ (e.g. by measuring them in a stable configuration). We then analyse two approaches. In the first one (FO) we apply a sequence of basis filter functions and reconstruct the signal by orthogonalization. The coefficients are obtained by comparing the reconstructed signal with the ansatz of Eq.~\eqref{eq:ansatz-composite-signal}. In the simulations we apply filters of $T=5$ duration and apply $K=10$ basis filter functions, leading to a total duration of $T_c=50$ for the measurement of one time instance of the coefficients $s_1$ and $s_2$ (we assume an instantaneous population measurement). We then repeat this procedure until we cover the time span we want to investigate.

In the second approach, we use the OCF protocol discussed above. We first apply the filter optimized for $S_1(\omega)$ and then the one for $S_2(\omega)$. Later, we use the measured overlap with the signal to calculate $s_1$ and $s_2$. Again, we use filter operation times of $T=5$, such that total duration is $T_c=10$ for one time instance. Thus it is $5$ times faster compared to the first approach. In the simulations we test it for $1$ and for $6$ qubits.

\begin{figure}
\centering
 \includegraphics[width=0.475\textwidth]{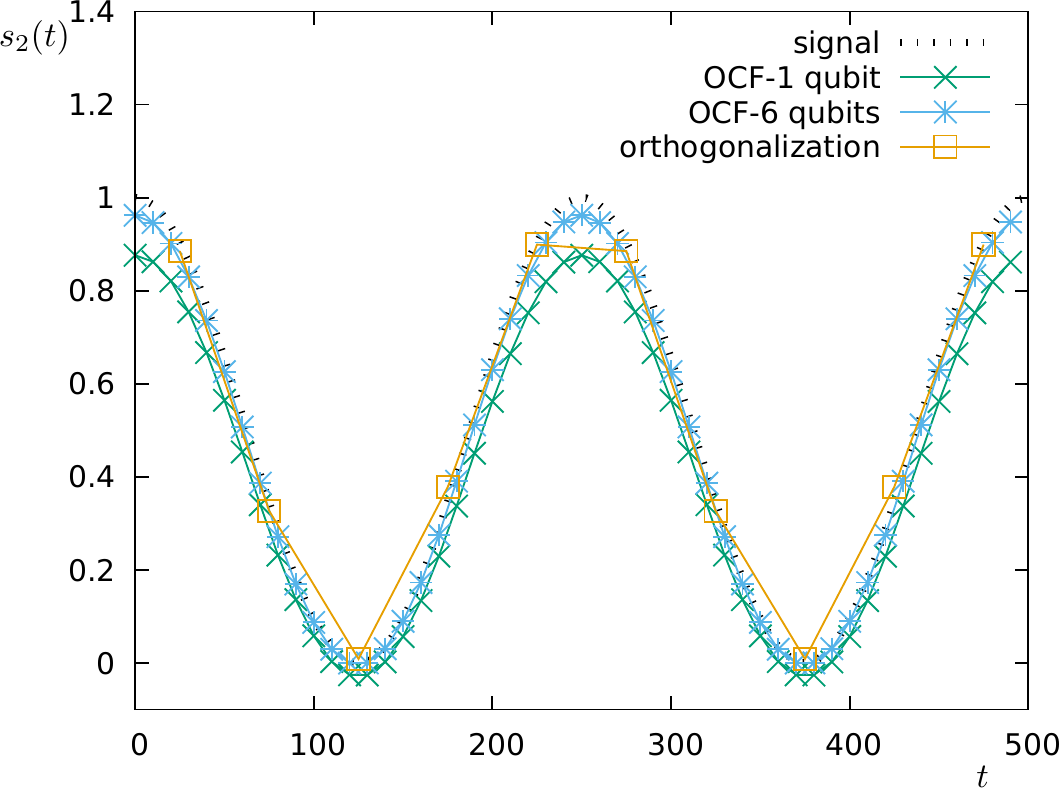}
 \caption{Detection of temporal changes in the spectrum. We plot the time-dependent coefficient $s_2(t)$ as modelled (signal: $s_2(t)=\cos^2(\omega_{osc} t)$, $\omega_{osc}=0.004\pi$) and as reconstructed by the two approaches FO (for 1 qubit) and OCF (for 1 and 6 qubits).}\label{fig:oscillation0004pi}
\end{figure}

\begin{figure}
\centering
 \includegraphics[width=0.475\textwidth]{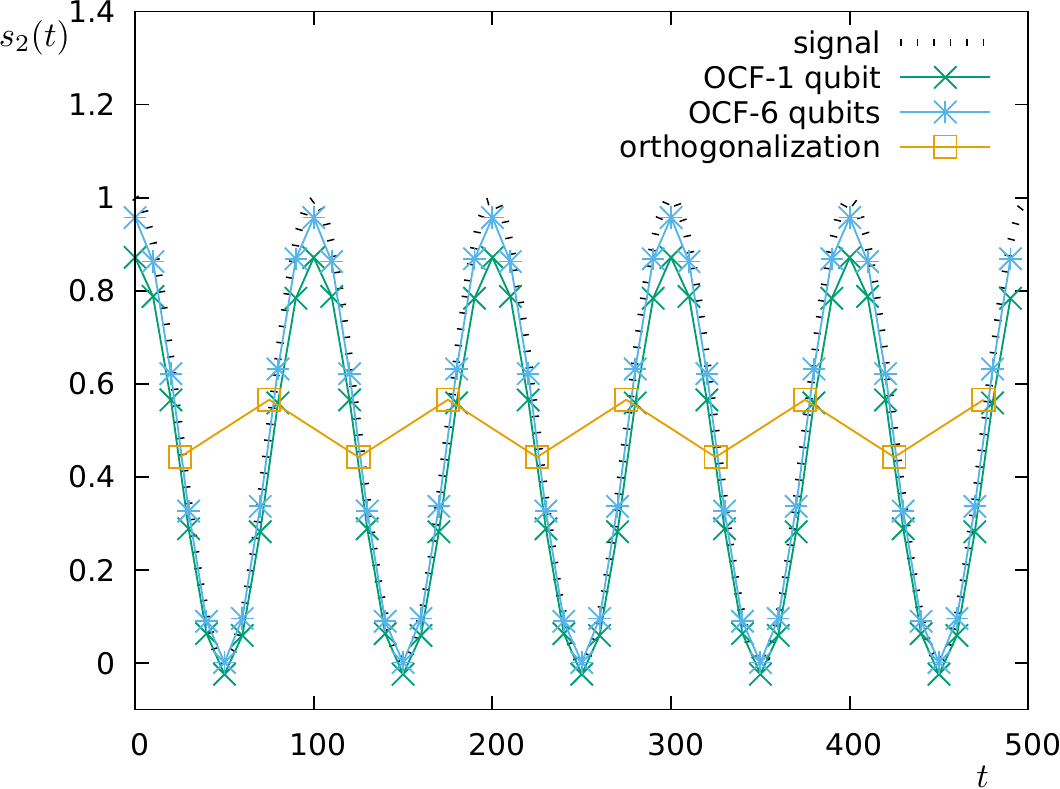}
 \caption{Detection of temporal changes in the spectrum. We plot the time-dependent coefficient $s_2(t)$ as modelled (signal: $s_2(t)=\cos^2(\omega_{osc} t)$, $\omega_{osc}=0.01\pi$) and as reconstructed by the two approaches FO (for 1 qubit) and OCF (for 1 and 6 qubits).}\label{fig:oscillation001pi}
\end{figure}

We show the results for two different values of $\omega_{osc}$ ($0.004\pi$ and $0.01\pi$) and analyse the time interval $t\in[0,500]$. The first approach can thus sample $10$ points, while the second approach $50$ points. Fig.~\ref{fig:oscillation0004pi} (for $\omega_{osc}=0.004\pi$) and Fig.~\ref{fig:oscillation001pi} (for $\omega_{osc}=0.01\pi$) show the oscillation of the coefficient $s_2(t)$ as measured via the two approaches. For the slower oscillation, the sampling rate of the first scheme (FO) is high enough to resolve the oscillation, and both schemes work well, while the second approach has a clear advantage when operated with 6 qubits. For the faster oscillation, however, the sampling rate of the first approach (FO) is too small, and the measured points do not correspond to the signal. The second approach (OCF), instead still works very well.

\section{Discussion}

We have introduced new concepts in quantum sensing of the power spectral density of a signal that allowed us to considerably speed up the sensing protocols and make them more robust against noise affecting the probe and the signal. In detail, we have introduced the concept of filter orthogonalization allowing us to use more arbitrary shapes of the filter functions and that might have future applications in the detection of multiple baths, where it could help to distinguish different single-bath contributions. We have studied a new type of error model that includes probe dephasing and finite detector precision and thus shown the robustness of the protocols in a realistic experimental situation as well as the advantage obtained by faster protocols. We have studied also statistical noise and its effect on the ultimate precision in terms of the Fisher information. In this regard, we have expanded the approach of \cite{Zwick2016}, where the signal is parametrized and the precision of the parameter measurement is optimized: by means of the concept of the Fisher Information Operator we provide a tool to optimize the detection of an arbitrary functional shape of the signal. We implement this approach by optimizing the filter overlap with a given signal. This extends optimal control theory to filter design in quantum noise spectroscopy and as an application we demonstrate that this new method provides a fast way to measure the contribution of two components of a composite signal. We have given two explicit examples of possible experimental realization, both for single qubits (with NV-centers in diamond) and entangled multi-qubits (with trapped ions), using parameters from existing experiments, showing thus, that an experimental realization of our new algorithms is feasible and advantageous with current technology and set-ups.
We have furthermore shown how multi-qubit probes and entanglement offer similar advantages as more complex control of the probe \cite{Frey2017} and we think that combining the two in a future work could provide new impulses to the study of networks of entangled quantum probes and the detection of both temporal and spatial resolutions \cite{Nokkola2016,Knott2016}.

\section{Methods}

\subsection{Estimation of the decoherence function}

The presented derivation follows roughly Refs.~\cite{KofmanNAT2000,KofmanPRL2001,Gordon}. We study a two-level system (with levels $|0\rangle$ and $|1\rangle$) governed by the Hamiltonian
\begin{eqnarray}
 H(t)=H_c(t)+H_S(t),\\
 H_c(t)=\frac{\Omega(t)}{2}\sigma_x\,,\qquad H_S(t)=E(t)\sigma_z\,,
\end{eqnarray}
with the control Hamiltonian $H_c$ and the signal Hamiltonian $H_S$, where $\Omega(t)$ is a control field that we can apply and $E(t)$ is a stochastic field stemming from an interaction with the environment. The Pauli operators are given by $\sigma_x=|0\rangle\langle 1|+|1\rangle\langle 0|$ and $\sigma_z=|1\rangle\langle 1|-|0\rangle\langle 0|$. The system is initially prepared in the state $|0\rangle$, which is also the state that we can measure. An initial $\pi/2$-pulse brings the population into the state $|\uparrow\rangle=\frac{|0\rangle + |1\rangle}{\sqrt{2}}$, which together with $|\downarrow\rangle=\frac{|0\rangle - |1\rangle}{\sqrt{2}}$ forms a basis of the system.
The dynamics of the system is then given by
\begin{equation}
 \dot{\rho}=-i[H_c,\rho]-i[H_S,\rho]\,,
\end{equation}
where $\rho$ is the density operator describing the state of the system. To simplify the calculations, we work in the rotating frame given by the control Hamiltonian. Namely, we consider the transformation
\begin{eqnarray}
 U_c(t)=e^{i\varphi(t)/2}|\uparrow\rangle\langle\uparrow| + e^{-i\varphi(t)/2}|\downarrow\rangle\langle\downarrow|\,,
\end{eqnarray}
where $\varphi(t)=\int_0^t\Omega(t')dt'$, and introduce the transformed state $\tilde{\rho}(t)=U_c(t)\rho(t) U_c^\dagger(t)$, as well as the transformed signal Hamiltonian $\tilde{H}_S(t)=U_c(t)H_S(t)U_c^\dagger(t)$. Then, we rewrite the Schr\"odinger equation as
\begin{equation}\label{eq:Schroedinger-transformed}
 \dot{\tilde\rho}(t)=-i[\tilde{H}_S(t),\tilde\rho(t)]\,.
\end{equation}
If we assume $\langle E(t)\rangle=0$ , i.e. if the stochastic field $E(t)$ on average vanishes, we can formally integrate the Schr\"odinger equation and reinsert $\tilde{\rho}(t)$ on the right hand side of Eq.~\eqref{eq:Schroedinger-transformed}, yielding
\begin{equation}
 \dot{\tilde\rho}=-\int_0^t \left\langle [\tilde{H}_S(t),[\tilde{H}_S(t'),\tilde\rho(t')] \right\rangle dt'\,.
\end{equation}
The master equation  has the fix point
\begin{eqnarray*}
\rho =
\begin{pmatrix} {\rho_{\uparrow\uparrow}} & {\rho_{\uparrow\downarrow}}\\
  {\rho_{\downarrow\uparrow}} & {\rho_{\downarrow\downarrow}}\end{pmatrix}
 =\begin{pmatrix} \frac{1}{2}&0\\0&\frac{1}{2}\end{pmatrix}\,,
\end{eqnarray*}
where we have expressed the density matrix in terms of its matrix elements in the rotating frame. The fix point is approached exponentially, so that the final (average) probability to find the system upon measurement in $|\uparrow\rangle$ (or $|0\rangle$, if we apply a $\pi/2$-pulse directly before the measurement) can be written as
\begin{equation}
p(t)=\frac{1}{2}\big(1-\mathrm{e}^{-\chi(t)}\big)\,.
\end{equation}
The decoherence function $\chi(t)$ is determined by the decay of the populations to the fix point.
By introducing the autocorrelation function $g(t-t')=\langle E(t)E(t')\rangle$ and after some algebra, we obtain
\begin{eqnarray}
 \dot{\rho_{\uparrow\uparrow}}(t)-\dot{\rho_{\downarrow\downarrow}}(t)=\qquad\\
 -4 \int_0^t g(t-t')\cos\left(\varphi(t)-\varphi(t')\right)
 ({\rho_{\uparrow\uparrow}}(t')-{\rho_{\downarrow\downarrow}}(t')) dt'\,.\nonumber
\end{eqnarray}
If we assume that the decay is slow compared to the correlation length given by $g(t-t')$, we can substitute $({\rho_{\uparrow\uparrow}}(t')-{\rho_{\downarrow\downarrow}}(t'))$ by $({\rho_{\uparrow\uparrow}}(t)-{\rho_{\downarrow\downarrow}}(t))$ and solve the equation. As a result, we obtain
\begin{equation}
 \chi(t)=4\int_0^t\int_0^t g(t'-t'')\cos(\varphi(t')-\varphi(t'')) dt'dt''\,.
\end{equation}
Moreover, if we introduce the power spectral density $S(\omega)$ through the Fourier transform
\begin{equation}
 g(\tau)=\frac{1}{2\pi}\int_{-\infty}^{\infty} S(\omega) \mathrm{e}^{i\omega\tau}d\omega\,,
\end{equation}
we get the decoherence function in the spectral form, i.e.
\begin{equation}
 \chi(t)=\frac{4}{\pi}\int_{0}^\infty\int_0^t\int_0^t S(\omega) \cos(\varphi(t')-\varphi(t'')) dt'dt''d\omega\,,
\end{equation}
where $\cos(\varphi(t'')-\varphi(t'))=y(t')y(t'')+z(t')z(t')$, with the pulse modulation functions $y(t)=\cos\varphi(t)$, $z(t)=\sin\varphi(t)$. If we introduce their Fourier transforms
\begin{eqnarray}
 Y(\omega)=\int_0^t y(t')e^{i\omega t'}dt'\,,\\
 Z(\omega)=\int_0^t z(t')e^{i\omega t'}dt'\,,
 \end{eqnarray}
and the Filter function $F(\omega)=\frac{4}{\pi}(|Y(\omega)|^2+|Z(\omega)|^2)$, the decoherence function becomes
\begin{equation}
 \chi(t)=\int_{0}^\infty S(\omega) F(\omega) d\omega\,.
\end{equation}

For the widely used control sequences consisting of a series of $\pi$-pulses initialized and ended by a $\pi/2$-pulse~\cite{Cywinski,Alvarez2011,Degen2016}, we have $\varphi(t)\in\{\pi,-\pi\}$ and thus $y(t)\in\{-1,1\}$ and $z(t)=0$.

\subsubsection{Entangled multi-qubit probes}

Here, we will derive the filter function for entangled multi-qubit probes and a control sequence consisting of single-qubit $\pi$-pulses (i.e. we assume single addressing of the qubits by the control knob and limit ourselves to $\pi$-pulses).
To explain the concept, let us initially consider two qubits and, instead of applying the initial (local) $\pi/2$-pulse, let us assume to prepare them in the Bell state $|\uparrow\rangle=\frac{|00\rangle + |11\rangle}{\sqrt{2}}$.

The sequence of the single-qubit $\pi$-pulses is a purely local transformation described by a transformation $U_c(t)$ that defines the rotating frame basis states $U_c(t)|\uparrow\rangle$ and $U_c(t)|\downarrow\rangle=U_c(t)\frac{|00\rangle-|11\rangle}{\sqrt{2}}$. Likewise, the fluctuating field $E(t)$ creates a purely local transformation, assuming that the field $E(t)$ is the same for both qubits, i.e. it is spatially homogeneous: it leads to an oscillation between $U_c(t)|\uparrow\rangle$ and $U_c(t)|\downarrow\rangle$. Here, we make these two assumptions (i.e. local transformations) and apply the filter function technique to a two-qubit system, with the modulation functions $y_1(t)$ and $y_2(t)$ for the first and second qubit, respectively.

If we write $y(t)=y_1(t)+y_2(t)$, the dynamics of the system in the two-dimensional rotating frame is described by
\begin{equation}
 \dot{\tilde\rho}=-iE(t)y(t) [X,\tilde\rho]\,,
\end{equation}
where $X=\begin{pmatrix}0&1\\1&0\end{pmatrix}$, $E(t)y(t)X$ is the signal Hamiltonian in the rotating frame, and $\tilde{\rho}$ is the two-dimensional density matrix describing the state in the effective two-dimensional rotating Hilbert space.
Assuming again $\langle E(t)\rangle=0$ and $g(t-t')=\langle E(t)E(t')\rangle$, we obtain
\begin{equation}
 \dot{\rho}=-2\int_0^t g(t-t')y(t)y(t')[X,[X,\rho(t')]]dt'
\end{equation}
and thus
\begin{equation}
 \dot{\rho_{\uparrow\uparrow}}(t)-\dot{\rho_{\downarrow\downarrow}}(t)=-2\int_0^t g(t-t')y(t)y(t')
 (\rho_{\uparrow\uparrow}(t')-\rho_{\downarrow\downarrow}(t')) dt'\,.
\end{equation}

Thus, the decoherence function in the frequency domain becomes
\begin{eqnarray}
 \chi(t)=\frac{4}{\pi}\int_0^{\infty} S(\omega) |Y(\omega)|^2 d\omega\,,
\end{eqnarray}
where this time the filter function $F(\omega)=|Y(\omega)|^2$ is defined by the Fourier transform of $y(t)=y_1(t)+y_2(t)$:
\begin{equation}
 Y(\omega)=\int_0^t y(t')e^{i\omega t'}dt'\,.
\end{equation}
In this case, the pulse modulation function takes the discrete values $y(t)\in\{-2,0,2\}$.

By generalizing this estimation procedure to a multi-qubits framework, we consider the GHZ state
\begin{equation}
\frac{|0...0\rangle + |1...1\rangle}{\sqrt{2}},
\end{equation}
with the modulation functions $y_i(t)$ for each of the single qubits and $y(t)\in\{-N,2-N,\dots, N-2, N\}$. The final filter function for $N$ qubits then reads
\begin{equation}
 F(\omega)=\frac{4}{\pi}|Y(\omega)|^2=\frac{4}{\pi}\bigg|\int_0^t \sum_{j=1}^N y_j(t') e^{i\omega t'}dt'\bigg|^2\,.\\
\end{equation}

\subsection{Fisher Information Operator}

The proposed protocols for the estimation of the power spectral density $S(\omega)$ rely on performing measurements of the survival probability $p(t)$ for different filter functions. The estimate $\tilde S(\omega)$ will generally deviate from $S(\omega)$ and we want to analyse how large such a deviation in a given direction $\overline{S}(\omega)$ in function space can be. If we assume that the main source of error is the statistical noise in the measurement of this survival probability and that on each realization of the measurement we obtain binary outcomes (survival or not), we can estimate the error $\varepsilon\overline{S}(\omega)$ (with $\varepsilon$ being a constant quantifying the deviation of $\tilde S(\omega)$ from $S(\omega)$ in the direction $\overline{S}(\omega)$) by the Fisher Information Operator (FIO) formalism as given in Ref.~\cite{MuellerFisher}. This is a straightforward generalization of the Fisher Information matrix to an (infinite dimensional) function space. The single FIO $\mathfrak{F}_{k}$, associated with the measurement of the $k-$th probability $p_{k}$ after the application of the filter function $F_{k}$, can be obtained by computing the derivative of $p_{k}$ with respect to $\chi_{k}(t)$ (decoherence function of the probability $p_{k}$):
\begin{equation}
 \frac{d p_{k}}{d\chi_{k}}(t)=\frac{1}{2}\mathrm{e}^{-\chi_{k}(t)}=1-p_{k}(t),
\end{equation}
as well as the (functional) derivative of $\chi_{k}(t)$ with respect to $S(\omega)$:
\begin{eqnarray}
 \bigg\langle\frac{\partial \chi_{k}}{\partial S}\bigg|(\cdot)\equiv\int_{0}^{\infty}(\cdot)F_{k}(\omega)d\omega.
\end{eqnarray}
This functional derivative can be understood by considering the variation
\begin{eqnarray}
 \delta \chi_{k}\equiv\int_{0}^{\infty}\delta S(\omega)F_{k}(\omega)d\omega.
\end{eqnarray}
Formally dividing by $\delta S(\omega)$ yields the functional derivative $\delta \chi_{k}/\delta S$ which is an element of the dual space of the tangent space of the spectral density functions, thus a linear mapping from the admissible changes $\delta S(\omega)$ to a real number $\delta\chi_k$. We then express this fact by the ket notation $\langle\cdot|$ as above.

By the chain rule, the FIO $\mathfrak{F}_{k}$ is equal to
\begin{eqnarray}
 \mathfrak{F}_{k} &=&\frac{1}{p_{k}(1-p_{k})}\left(\frac{d p_{k}}{d\chi_{k}}\right)^2\bigg|\frac{\partial \chi_{k}}{\partial S}\bigg\rangle \bigg\langle\frac{\partial\chi_{k}}{\partial S}\bigg|\\
  &=&\frac{1-p_{k}}{p_{k}}\bigg|\frac{\partial\chi_{k}}{\partial S}\bigg\rangle\bigg\langle\frac{\partial\chi_{k}}{\partial S}\bigg|.
\end{eqnarray}
Finally, if we perform a sequence of measurements of $p(t)$ by using several filter functions $F_k$, $k=1,\dots, K$, then the overall FIO $\mathfrak{F}$ (for the sequence of measurements) is additive, since the measurements are statistically independent, and, as a result, we get
\begin{equation}
 \mathfrak{F} = \sum_{k=1}^K \frac{1-p_k}{p_k}\bigg|\frac{\partial \chi_k}{\partial S}\bigg\rangle \bigg\langle\frac{\partial \chi_k}{\partial S}\bigg|.
\end{equation}

The rank of this FIO $\mathfrak{F}$ corresponds to the number of linear independent vectors $\big\langle\frac{\partial \chi_k}{\partial S}\big|$, which equals to the number of linear independent filter functions $F_k$ adopted for the estimation of $S(\omega)$. Thus, for an adequate choice of the pulse sequences (i.e. of the filter functions) the rank of the FIO is $K$. This means, that using such $K$ filter functions we can determine the behaviour of the spectrum in a $K$-dimensional subspace of the function space of the spectral densities. This is a generalization of the common approach (e.g. for the AS protocol), where $K$ filter functions are used to determine the value of the spectrum at $K$ discrete values of the frequency. The behaviour of the spectrum outside of this subspace (or in the discrete case: for any other value of the frequency) is completely inaccessible by the given choice of filter functions.

\subsubsection{Cramér-Rao bound}

A bound for the error $\varepsilon$ is given by the Cramér-Rao bound associated with the Fisher Information for the parameter estimation of $\varepsilon$.
As shown in Ref.~\cite{MuellerFisher}, this Fisher information is obtained from the FIO as
\begin{eqnarray}
 \mathfrak{F}_{\overline{S}}&=&\left\langle\overline{S}|\mathfrak{F}|\overline{S}\right\rangle\\
 &=&\sum_{k=1}^K \frac{1-p_k}{p_k}\left(\int_{0}^\infty \overline{S}(\omega) F_k(\omega)d\omega\right)^2\,,
\end{eqnarray}
where $\overline{S}(\omega)$ is, as before, a given direction in the function space of the spectral densities.
The Cramér-Rao bound for $\varepsilon$ is then given by
\begin{equation}
 \varepsilon\geq \frac{1}{\sqrt{\mathfrak{F}_{\overline{S}}}}\,.
\end{equation}
Note that this bound is finite iff $\overline{S}(\omega)$ has finite overlap with at least one filter function $F_k(\omega)$. This corresponds to the fact that the application of $K$ (linearly independent) filters determines the signal $S(\omega)$ in  a K-dimensional subspace of the full function space.

\section{Acknowledgements}
The authors gratefully acknowledge Gershon Kurizki, J\"{o}rg Wrachtrup and Durga Dasari for useful discussions. This work was financially supported from the Fondazione CR Firenze through the project Q-BIOSCAN.

\end{document}